\documentclass[twocolumn]{aa}

\usepackage{graphicx}
\usepackage{amsmath,amssymb}
\usepackage{txfonts}
\usepackage[figuresright]{rotating}
\usepackage{natbib}
\bibpunct{(}{)}{;}{a}{}{,} 
\usepackage[dvipdfm,breaklinks=true,bookmarks=true,hypertexnames=false,bookmarksnumbered=true,bookmarksopen=false,bookmarks=true,bookmarksnumbered=true]{hyperref}

\newcommand{\msun}{{\rm M}_{\odot}}
\newcommand{\kms}{\, {\rm km\, s}^{-1}}
\newcommand{\h}{\,h_{\rm 70}}

\newcommand{\Mpc}{\, {\rm Mpc}}

\newcommand{\der}{{\rm d}}

\newcommand{\reffig}[1]{Fig.~\ref{#1}}

\newcommand{\mslsun}{\h\,(M/L)_\odot}

\newcommand{\mypm}[2]{^{+#1}_{-#2}}

\newcommand{\scrit}{\Sigma_{\rm crit}}
\newcommand{\dos}{D_{\rm os}}
\newcommand{\dol}{D_{\rm ol}}
\newcommand{\dls}{D_{\rm ls}}
\newcommand{\dlsdos}{\dls/\dos}

\newcommand{\zphot}{z_{\rm phot}}

\newcommand{\coordra}[3]{#1^{\rm h}~#2^{\rm m}~#3^{\rm s}}
\newcommand{\coorddec}[3]{#1^{\rm o}~#2{\rm '}~#3{\rm ''}}

\newcommand{\eg}{{\it e.g.}~}
\newcommand{\ie}{{\it i.e.}~}

\newcommand{\nbg}{n_{\rm bg}}

\newcommand{\zl}{z_{\rm l}}
\newcommand{\zs}{z_{\rm s}}

\newcommand{\ugriz}{$u^*g'r'i'z'$}
\newcommand{\rein}{\theta_{\rm E}}

\begin{document}

\title{Weak lensing survey of galaxy clusters in the CFHTLS Deep\thanks{
    Based on observations obtained with MegaPrime/MegaCam, a
    joint project of CFHT and CEA/DAPNIA, at the Canada-France-Hawaii
    Telescope (CFHT) which is operated by the National Research
    Council (NRC) of Canada, the Institut National des Science de
    l'Univers of the Centre National de la Recherche Scientifique
    (CNRS) of France, and the University of Hawaii. This work is based
    in part on data products produced at TERAPIX and the Canadian
    Astronomy Data Centre as part of the Canada-France-Hawaii
    Telescope Legacy Survey, a collaborative project of NRC and
    CNRS.}
}
\titlerunning{Weak lensing in CFHTLS Deep fields}

\author{R. Gavazzi\inst{1} \and  G. Soucail\inst{1} }
\authorrunning{Gavazzi \& Soucail}

\offprints{Rapha\"el Gavazzi, \email{gavazzi@physics.ucsb.edu}}

\institute{
  Laboratoire d'Astrophysique de Toulouse-Tarbes, 
  Observatoire Midi-Pyr\'en\'ees,
  UMR5572 CNRS \& Universit\'e Paul Sabatier,
  14 Avenue Edouard Belin, 31400 Toulouse, France
}

\date{Received sometime; accepted later}

\abstract
{}
{
  We present a weak lensing search of galaxy clusters in the 4 deg$^2$
  of the CFHT Legacy Survey Deep. This work aims at building a
  mass-selected sample of clusters with well controlled selection
  effects. This present survey is a preliminary step toward a full
  implementation in the forthcoming 170 deg$^2$ of the CFHTLS Wide
  survey.
} 
{
  We use the deep $i'$ band images observed under subarcsecond seeing
  conditions to perform weak lensing mass reconstructions and to
  identify high convergence peaks. Thanks to the availability of deep
  \ugriz~exposures, sources are selected from their photometric
  redshifts in the weak lensing analysis. We also use lensing
  tomography to derive an estimate of the lens redshift. After
  considering the raw statistics of peaks we check whether they
  can be associated to a clear optical counterpart or to published
  X-ray selected clusters.
} 
{
  Among the 14 peaks found above a signal-to-noise detection threshold
  $\nu=3.5$, nine are secure detections with estimated redshift
  $0.1\lesssim \zl\lesssim0.7$ and a velocity dispersion
  $450\lesssim\sigma_v\lesssim 700 \kms$. This low mass range is
  accessible thanks to the high density of background sources.
  Considering the intersection between the shear-selected
  clusters and XMM-LSS X-ray clusters in the D1 field, we observe that
  the ICM gas in these low-mass clusters ($T_X\sim1-2$ keV) is not
  hotter than the temperature inferred from shear, this trend being
  different for published massive clusters. A more extended weak
  lensing survey, with higher statistics of mass structures will be a
  promising way to bypass several of the problems related to standard
  detection methods based on the complex physics of baryons.
} 
{}


\maketitle

\section{Introduction}\label{sec:intro}
In the context of hierarchical structure formation within the Lambda
Cold Dark Matter paradigm ($\Lambda$CDM), clusters of galaxies are the
very latest structures to assemble. They are also the most extended
gravitationally bound systems in the Universe and constitute a key
laboratory for cosmology. The time evolution of clusters as well as
their abundance and spatial clustering properties are essentially
driven by gravity sourced by the CDM mass content
\citep[\eg][]{eke96b,haiman01,borgani01}.

Therefore the main physical parameter for a cluster of galaxies is its
total mass. However most of observations only have an indirect access
to cluster masses and rather measure quantities like the SZ decrement,
X-ray or optical/NIR luminosities, the line-of-sight velocity
dispersion of the cluster-member galaxies or the temperature of the
hot gas in the intra-cluster medium (ICM) from X-ray
\citep[\eg][]{bahcall95,carlberg96,bahcall03,olsen99b,gladders00,romer01,carlstrom02}.
Hence difficulties arise because one needs well calibrated proxies to
convert observables into theoretically relevant quantities like mass
and because the process of detecting clusters of galaxies with such
indirect methods might suffer various kinds of selection effects.
Therefore any attempt to use clusters of galaxies as efficient
cosmological probes cannot afford making extensive assessments of
the assumed calibration of the scaling laws in the local Universe
and at high redshift \citep[\eg][]{arnaud99,dasilva04,arnaud05}. 

Gravitational lensing is among the best ways to test biases in the
above techniques. The bending of light by intervening matter along
the line of sight from distant sources to the observer only depends
on the mass properties of structures without regards of its nature
(baryonic or not, luminous or dark) or dynamical state (relaxed or
not, hydrostatic equilibrium...). Since the early 90's several groups
have reported the detection of a weak lensing signal around massive
clusters of galaxies. However the broad range of observational
configurations (field of view, depth, seeing, ground- or space-based
images, etc) makes difficult a direct comparison of published
results. For reviews see \citet{mellier99} and
\citet{BS01}. Progresses have been made in this direction with weak
lensing studies of sizable samples of optically or X-ray selected
clusters \citep{dahle02,cypriano04,clowe06,bardeau06}. Different mass
estimates globally agree although outliers perturb a simple relation
between X-ray (or dynamical) and lensing mass estimates
\citep[\eg][]{allen98,wu00,arabadjis04}. This suggests that dynamical
activity is still important for massive halos and that asphericity and
projection effects may complicate both weak lensing and other mass
estimates
\citep{metzler01,hoekstra03,clowe04,deputter05,gavazzi05,filippis05}. 

In parallel, the idea of direct detection of galaxy clusters by their
weak lensing signal starts to emerge. By measuring the coherent
stretching of distant galaxies by intervening structures, one is able
to infer the projected density field (\ie the so-called
convergence). Hence high convergence peaks may be identified as
massive clusters of galaxies. This is the idea of a direct weak
lensing cluster survey (hereafter WLCS), aimed at building a
mass-selected cluster sample directly comparable to CDM theory
(through N-body cosmological simulations).  On the theoretical side,
pioneering analytical predictions based on the mass function of halos
have been proposed \citep{schneider96,kruse99}, but they were not able
to properly account for projection effects \citep{reblinsky99}. In
addition \citet{bartelmann01} showed that WLCSs are very sensitive to
the details of the clusters density profile. More recently, ray-tracing
into N-body cosmological simulations have been used to properly
address the critical issue of projections and clusters' asphericity
\citep{white02,padmanabhan03,hamana04,tang05} and the way
to reduce the effect of noise on cluster detections through an optimised
data filtering procedure
\citep{hennawi05,maturi05,starck06}. A step forward will
also probably be made with simplified analytical models
of the convergence one-point PDF \citep{taruya02,das05}.
Under standard observational conditions, these works predict that at
$z\sim0.2$ clusters more massive than $M\sim5\times 10^{13}\msun$ can be
recovered with a signal-to-noise ratio $\nu=3$. This limit drops to
$M\sim2\times 10^{14}\msun$ at $z\sim0.7$ . Therefore the main targets
of WLCSs are massive clusters of galaxies.

From the point of view of observations, we can mention a few
serendipitous detections of galaxies clusters via weak gravitational
lensing \citep[\eg][]{schirmer03,schirmer04,dahle03b}. A few examples
have also been found to show up through weak lensing techniques
without any clear optical counterpart and gave support for the existence
for the so-called {\it ``dark clumps''} \citep{umetsu00,miralles02}.
The practical implementation of a systematic WLCS is however very
new. \citet{miyazaki02} studied an area of 2.1 deg$^2$ with
Suprime-Cam on Subaru telescope under excellent seeing
conditions. They report an excess of $4.9\pm2.3$ convergence peaks
with signal-to-noise $\nu>5$. \citet{hetterscheidt05} report the
detection of $\sim 5$ cluster candidates over a set of 50 disconnected
VLT/FORS deep images covering an effective area of 0.64 deg$^2$ while
\citet{wittman06} present preliminary results for the first 8.6
deg$^2$ of the Deep Lens Survey (eight detections).  \citet{haiman04}
also make interesting predictions for future WLCS applications in the
LSST survey.

In this paper we present a weak lensing analysis of the Deep CFHT
Legacy Survey\footnote{\url{http://www.cfht.hawaii.edu/Science/CFHTLS/}}
covering 4 $\times$ 1 deg$^2$ in five optical bands (\ugriz) under subarcsec
seeing condition as a pilot analysis for the ongoing Wide Survey which
will cover 170 deg$^2$. The present work proposes to carry out weak
lensing mass reconstructions in the Deep fields and focus on high
convergence peaks in order to shed light on WLCS capabilities. The
relatively high sample variance of the Deep images prevents any
cosmological application of WLCSs but the great depth and amount of
photometry make us with an excellent laboratory for a future
application to the Wide data.

The paper is organised as follows. In Sect. \ref{sec:lensdef} we
briefly review the basics of weak gravitational lensing. In
Sect. \ref{sec:data} we present the data at hand, the specific
treatment required for weak lensing signal extraction and photometric
redshifts. We show mass reconstructions (\ie convergence maps) in
Sect. \ref{sec:massmap} inferred from the coherent shear field
imprinted on distant background sources. We also measure the statistics
of high convergence peaks whereas we focus on their properties
in Sect. \ref{sec:peaks} by studying the associated optical and X-ray
counterparts (when available). We discuss our main achievements
and conclude in Sect. \ref{sec:conclu}.

In the following we assume the {\it ``concordance model''}
cosmological background with $H_0 = 70\,\kms\Mpc^{-1}$,
$\Omega_{\rm m}=0.3$ and $\Omega_\Lambda=0.7$ . All magnitudes are
expressed in the AB photometric system.

\section{Basic lensing equations}\label{sec:lensdef}
In this section we briefly summarise the necessary background of
gravitational lensing and especially the weak lensing regime which
concerns the present analysis. We refer the reader to the reviews of
\citet{mellier99} and \citet{BS01} for more detailed accounts.

The fundamental quantity for gravitational lensing is the lens
Newtonian potential $\psi(\vec{\theta})$ at angular position
$\vec{\theta}$ which is related to the surface mass density
$\Sigma(\vec{\theta})$ projected onto the lens plane through
\begin{equation}
\psi(\vec{\theta}) = \frac{4 G}{c^2} \frac{\dol\dls}{\dos} 
\int \der^2 \theta' \Sigma(\vec{\theta}') \ln\vert \vec{\theta}-\vec{\theta}'\vert\;,
\end{equation}
where $\dol$, $\dos$ and $\dls$ are angular distances to the lens,
to the source and between the lens and the source respectively.
The deflection angle $\vec{\alpha}=\vec{\nabla}\psi$ relates a point
in the source plane $\vec{\beta}$ to its image(s) in the image plane
$\vec{\theta}$ through the lens equation
$\vec{\beta}=\vec{\theta}-\vec{\alpha}(\vec{\theta})$.
The local relation between $\vec{\beta}$ and $\vec{\theta}$
is the Jacobian matrix $a_{ij}=\partial \beta_i/\partial \theta_j$
\begin{equation}\label{eq:jacob}
  a_{ij} = \delta_{ij} - \psi_{,ij} = \left(\begin{array}{cc}
      1-\kappa-\gamma_1 & -\gamma_2 \\
      -\gamma_2  & 1-\kappa+\gamma_1
    \end{array}\right)\;.
\end{equation}
The convergence $\kappa(\vec{\theta})=\Sigma(\vec{\theta})/\scrit$ 
is directly related to the surface mass density via the critical density
\begin{equation}\label{eq:scrit}
  \scrit = \frac{c^2}{4 \pi G}\frac{\dos}{\dol\dls}\;,
\end{equation}
and satisfies the Poisson equation
\begin{equation}\label{eq:poisson}
\Delta \psi = \psi_{,11}+\psi_{,22} = 2 \kappa 
\end{equation}
The 2-component shear is
$\gamma=\gamma_1+i \gamma_2 = \frac{1}{2}(\psi_{,11}-\psi_{,22})+i\psi_{,12}$
in complex notation. An elliptical object in the image plane is
characterised by its complex ellipticity $e$ defined from the second
moments tensor $Q_{ij}$ of the surface brightness distribution as
\begin{equation}\label{eq:ellipticity}
e = \frac{Q_{11} - Q_{22} + 2i Q_{12}}{Q_{11} + Q_{22} + 2 \
\left(Q_{11}Q_{22} - Q_{12}^2\right)^{1/2}}
\end{equation}
This observed ellipticity can be related to the intrinsic ellipticity
of the source $e_s$ by:
\begin{equation}\label{eq:sheardef}
  e = \left\{\begin{array}{cc}
        \displaystyle 
        \frac{e_s + g}{1+g^* e_s} & {\rm , for}~~\lvert g\rvert\le 1\\
\\
        \displaystyle
        \frac{1 + g e_s^*}{e_s^* + g^*} & {\rm , for}~~\lvert g\rvert > 1
      \end{array}\right.
\end{equation} 
where $g=\gamma/(1-\kappa)$ is the so-called reduced shear and $*$
denotes complex conjugation \citep{seitz97,geiger98}. In the weak
lensing regime ($g\sim\gamma \ll 1$), Eq. \eqref{eq:sheardef} reduces
to $e=e_s+\gamma$. Provided the random orientation of sources
reduces the averaged source ellipticity to zero, $e$ provides an
unbiased estimate for the shear $\gamma$. The noise associated to this
estimator is due to the scatter in the intrinsic ellipticity of
sources with a typical value $\sigma_e\sim0.3$ per component.

In the equations above we can isolate a geometric term which linearly
scales the lensing quantities $\kappa$, $\psi$, and $\gamma$ and only
depends on the distance ratio $\dlsdos$. We thus can write
$\gamma=w(\zl,\zs) \gamma_\infty$ (and so forth for $\kappa$ and
$\psi$) with $w(\zl,\zs)=\dlsdos \Theta(\zs-\zl)$ and $\Theta(x)$ is
the Heavyside step function. If sources are not confined in a thin plane,
we account for the distribution in redshift by defining an ensemble
average distance factor $W(\zl)$ such that:
\begin{equation}\label{eq:zweight}
  W(\zl)=\langle w(\zl,\zs)\rangle_{\zs}= \frac{\displaystyle \int_{\zl}^\infty
    \der \zs\, n(\zs) \frac{\dls}{\dos}}
  {\displaystyle\int_0^\infty  \der \zs\, n(\zs)}\,.
\end{equation}

If we now consider a broad distribution of mass intervening between
sources and the observer, we can use the Born approximation for the
propagation of light in a clumpy Universe to infer the effective
convergence experienced by light bundles:
\begin{equation}\label{eq:kappaeff}
  \begin{split}
  \kappa_{\rm eff}(\vec{\theta})= &
  \int_0^{l_H}\frac{\rho(l,\vec{\theta}) -
    \overline{\rho}(l)}{\scrit(l)} \, \der l\\
  = &\frac{3}{2} \Omega_m \left(\frac{H_0}{c}\right)^2
  \int_0^{\chi_H} \der \chi\, W(\chi) f_k(\chi)
  \frac{\delta( f_k(\chi)\vec{\theta},\chi)}{a(\chi)}\,,
  \end{split}
\end{equation}
with $l$ (resp. $\chi$) the proper (resp. comoving) distance,
$a$ the scale factor and $\delta(\vec{r})$ the density contrast.
At this level we clearly see the projection nature of weak lensing
signal. This means that a given $\kappa$-peak can be the result of a
single  $\delta$-peak (\eg a cluster of galaxies) or the sum of two or
more less pronounced $\delta$-peaks (\eg groups or filaments aligned
along the line of sight...).

\begin{table*}[htp]
  \centering
  \caption{Summary of CFHTLS Deep data, release T0003
    (T0002/T0003 for $i'$ band). The limiting magnitudes correspond
    to 50\% completeness and are expressed in AB system.
    Seeing is measured in T0002 release $i'$ images.}\label{tab:data-summary}
  \begin{tabular}{lcccc}\hline\hline
    & D1 & D2 & D3 & D4\\ \hline
    $\alpha_{2000}$  & $\coordra{02}{25}{59}$& $\coordra{10}{00}{28}$& $\coordra{14}{19}{27}$& $\coordra{22}{15}{31}$\\ 
    $\delta_{2000}$ & $\coorddec{-04}{29}{40}$ & $\coorddec{+02}{12}{30}$& $\coorddec{+52}{40}{56}$&$\coorddec{-17}{43}{56}$\\
    $z'$   & 25.0 & 24.9 & 25.1 & 25.0\\
    $i'$   & 25.9/25.9 & 25.4/25.7 & 25.9/26.2 & 25.7/26.0 \\
    $r'$   & 25.0 & 26.0 & 26.4 & 26.4 \\
    $g'$   & 26.4 & 26.2 & 26.6 & 26.3 \\
    $u^*$  & 26.5 & 26.1 & 25.9 & 26.5 \\
    $i'$ seeing ($\arcsec$)  & 0.91& 0.95& 0.91&0.87\\
    Area (deg$^2$) & 0.93& 0.89& 0.92&0.86 \\
    \hline
  \end{tabular}
\end{table*}

\section{The data}\label{sec:data}
\subsection{Imaging and Photometry}\label{ssec:data:desc}
The CFHTLS Deep survey is intimately linked to the Supernovae Legacy Survey
\citep[SNLS,][]{astier06} as they share the same data. In practice,
for each observed field data are taken sequentially every 3-4 nights, 6
months per year and in 4 observing bands ($g', r', i', z'$).
Additional $u^*$
data are included but they are not part of the SNLS and do not require
time sampling. Most of the data have a seeing requirement limited to
0.9\arcsec . Data acquisition is still under progress at CFHT so the depth
of the Deep fields is still improving.  Data processing (astrometry,
photometric calibration, final stacking and production of catalogues)
is performed at Terapix\footnote{\url{http://terapix.iap.fr/}} for
the CFHTLS community. The final data are released regularly by the
CADC\footnote{\url{http://cadcwww.hia.nrc.ca/}}. The present analysis
is based on 2 sets of data released subsequently (the details of the
release contents can be found on the Terapix Web site). For the weak
lensing signal extraction we used the T0002 data while the deeper images
and catalogues from the T0003 release were included for the photometric
redshift production. The shear analysis (see below) was already
performed on the T0002 data release at the time of T0003 data release.
Since the $i'$ band images did not gain much depth between T0002 and T0003,
the weak lensing analysis, which is based on shape measurements of galaxies
and is not sensitive to magnitude completeness, remained almost unchanged.
Conversely the gain in exposure time and limiting magnitudes for other
filters was extremely useful for photometric redshifts
(see Sect. \ref{ssec:data:photoz}).

\begin{figure*}[tbh]
  \centering
  \includegraphics[width=16cm]{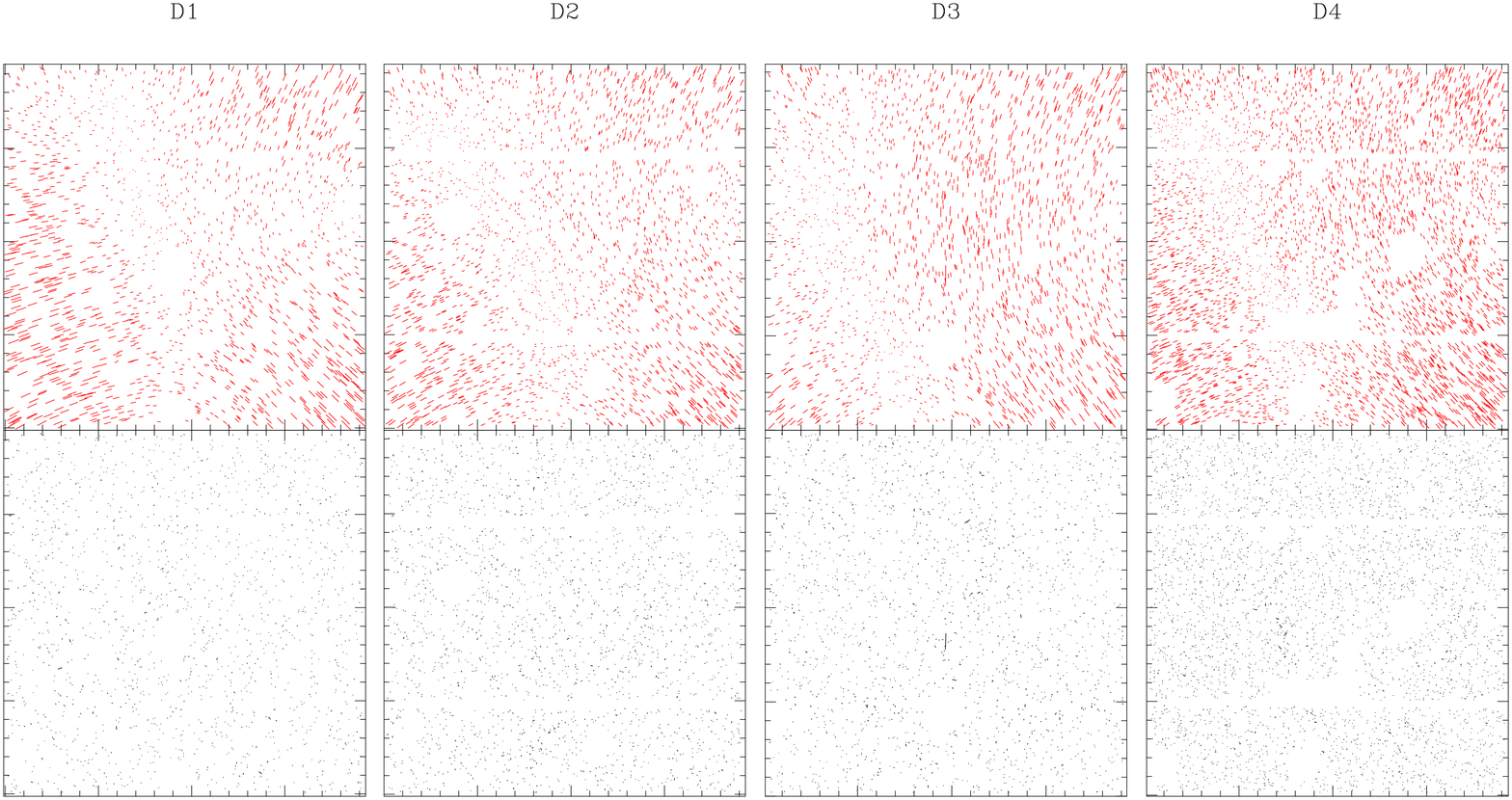}
  \includegraphics[width=16cm]{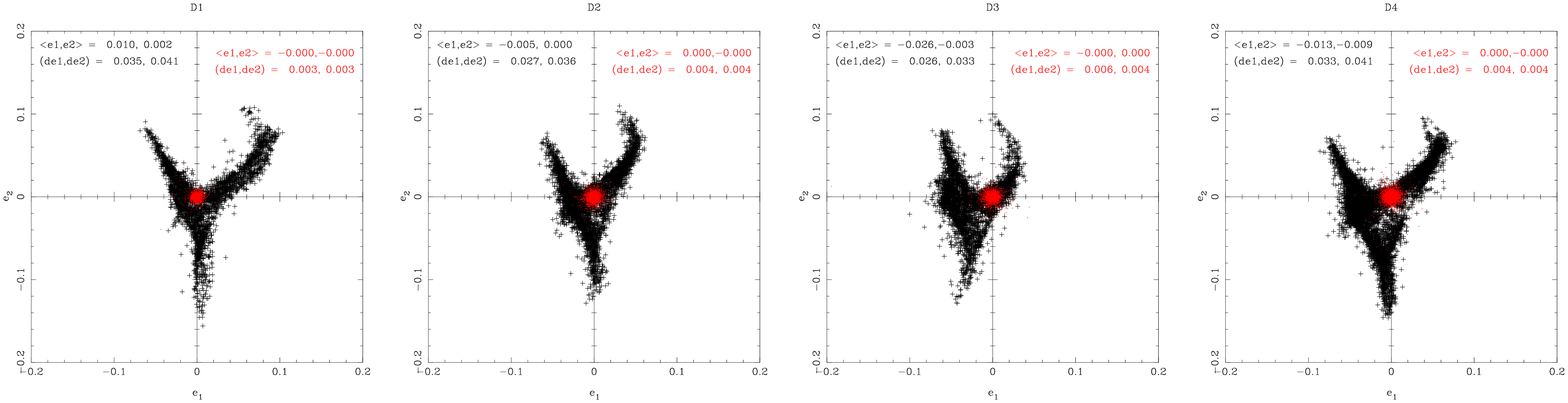}
  \caption{\small Stellar ellipticities in the CFHTLS Deep fields for
    D1... D4 from left to right. {\it Upper row:} polarisation field
    before and after PSF anisotropy correction (respectively red upper
    and lower black panels.). {\it Lower row:} projection of the
    stellar ellipticities in the $(e_1,e_2)$ plane before and after
    PSF anisotropy correction (black crosses and red dots
    respectively).}
  \label{fig:psf-anis}
\end{figure*}

The Deep survey is made of four independent patches called D1, D2, D3 and D4.
For each field and filter, Table \ref{tab:data-summary} summarises the
main observational properties of the T0002/T0003 release data
in terms of coordinates, seeing, exposure time and depth. Because of
the presence of bright stars, fields boundaries, defects in the CCDs and
gaps between them with low signal-to-noise ratios, a substantial part of
the images cannot be used for weak lensing analysis. Hence, the masked
regions generally result in a loss of 20\% of the field area.
In \reffig{fig:psf-anis}, holes in the distribution of stars reveal
the masked regions. The effective usable area is given in Table
\ref{tab:data-summary}. The total working area for the weak lensing
analysis is 3.61 deg$^2$.

\subsection{Shear measurement}\label{ssec:shear-meas}
The coherent stretching of background sources produced by the weak
lensing effect is measured using $i'$ band images. From one
field to another the depth of the catalogues is slightly varying but
for better coherence between the fields, we define a common 
magnitude cut for the selection of background sources $22<i'<26$. In
addition, close galaxy pairs with angular separations less than
3\arcsec\ are discarded to avoid blended systems with biased ellipticity
measurements. Small galaxies with a half flux radius $r_h$ smaller
than that of stars are also rejected.

The reliability of shear measurements is expected to be comparable to
the current cosmic shear survey analyses like the one already performed
by \citet{semboloni06} and \citet{hoekstra06} who found similar results
with independent pipelines in the same fields.
Throughout this paper, we report results of shear analysis performed
in the $i'$ band which presents the best balance between depth and seeing.
However we checked that we can extract a similar signal in the other noisier
$g', r'$ and $z'$ bands. This has also been assessed by \citet{semboloni06}
who report a similar cosmic shear signal in the deep fields using both
$r'$ and $i'$ bands.

Blurring and distortion of stars and galaxies produced by instrument
defects, optical aberration, telescope guiding, atmospheric seeing and
differential refraction are corrected using the Point Spread Function
(PSF) of stars over the whole field. Several correction techniques and
control of systematic errors have been proposed over the past 10 years
\citep[see \eg][]{mellier99,BS01}. In the following we use the most
popular KSB method initially proposed by \citet{KSB95}. Several teams
have already demonstrated that this technique can correct systematics
down to a lower value than the very weak cosmic shear signal
\citep{waerbeke05,heymans06step}. It is therefore well suited for this
analysis too. The KSB implementation used here is identical to that of
\citet{gavazzi04}.

The observed ellipticity components $ e^{\rm obs}_{\alpha=1,2}$ are
made of the intrinsic ellipticity components $e^{\rm src}_\alpha$, and
linear distortion terms that express the instrument and atmospheric
contaminations and the contribution of gravitational shear to the
galaxy ellipticity. Each ellipticity component is transformed as:
\begin{subequations}\label{eq:KSB}
  \begin{gather}
    e^{\rm obs}_\alpha = e^{\rm src}_\alpha + P^\gamma_{\alpha\beta} g_\beta +
    P^{\rm sm}_{\alpha\beta}q^*_\beta,\label{eq:KSBa}\\
    {\rm with\ \ } P^\gamma_{\alpha\beta} = P^{\rm sh}_{\alpha\beta}
    - P^{\rm sm}_{\alpha\gamma} \left(\frac{P^{\rm sh}}{P^{\rm sm}}\right)_
    {\gamma\beta}^*,\label{eq:KSBb}
  \end{gather}
\end{subequations}
where $g$ is the reduced gravitational shear, $P^{\rm sm}$ is the
{\sl smear polarisability}, $P^{\rm sh}$ the {\sl shear polarisability}
and $P^\gamma$ the isotropic circularisation contribution to the final
smearing. Sources are detected with SExtractor \citep{bertin96} but
shape parameters are calculated with {\tt
  Imcat}\footnote{\url{http://www.ifa.hawaii.edu/~kaiser/imcat/}}.
Because the noise present in these measured quantities is important,
all these tensors are simplified to half their trace
\citep[see \eg][]{erben01}.

$\left(\frac{P^{\rm sh}}{P^{\rm sm}}\right)^*$ and $q^*$ are measured
from field stars. Their spatial variation across the field is fitted
by a second order polynomial, applied individually to each one of the
36 CCDs composing the MegaCam focal plane. Stars are selected in
the magnitude-$r_h$ plane, as usual. $q^*$ is the anisotropic part of
the PSF, which is subtracted from observed ellipticities. The residuals
for stars are shown in \reffig{fig:psf-anis}~. After correction, these
latter are consistent with a $\sigma_\gamma\simeq0.004$ rms
featureless white noise. In Sect. \ref{sec:massmap1}, we present mass
reconstructions inferred from the shear field measured on distant
source galaxies. If we perform the same reconstructions on stars,
we only get white noise
as expected from a correct PSF smearing correction. Its rms is
$\sigma_\kappa\sim0.001$ when smoothed with a Gaussian kernel of width
1 arcmin, which is much below the signal we are interested in
(typically $0.01\lesssim\kappa\lesssim0.5$, see below).

The smearing part of the PSF contained in the $P^\gamma$ term depends
on the magnitude of the object and on its size as compared to the seeing disk.
To optimally extract $P^\gamma$, we derived it from an averaged value over its
40 nearest neighbours in the magnitude$-r_h$ plane. The variance of
ellipticities inside this neighbourhood is then used as a weighting
scheme for the shear analysis. The weight assigned to each galaxy is :
\begin{equation}\label{eq:weight_scheme}
  w_i = \frac{1}{\sigma_{e,i}^2} = \frac{P^{\gamma\,2}}
  {P^{\gamma\,2}\sigma_0^2 + \sigma_i^2},
\end{equation}
where $\sigma_0$ prevents from over-weighting some objects. It is set to 
the 1D intrinsic dispersion in galaxy ellipticities $\sigma_e=0.23$ and
$\sigma_i$ is the observed dispersion of ellipticities over the 40 nearest
neighbours in the magnitude-$r_h$ plane.

At this level, we have 132,000 (resp. 104,000, 162,000 and 114,000)
galaxies with reliable shape parameters in the D1 (resp. D2, D3 and D4)
field leading to a source surface number density of $\nbg=38.0$
(resp. 30.6, 35.4, 34.5) arcmin$^{-2}$. These values are much higher
than the usual ones in weak lensing studies, which turn around 15-20
galaxies arcmin$^{-2}$. The magnitude cut $i'<26$
explains these high densities although it makes difficult an
accurate determination of the redshift distribution of such faint
objects. In the following we shall refer to this source catalogue
as CA. Because several galaxies have large uncertainties on their
shape parameters, one should correct this density by considering the
effective density which would have been achieved if all uncertainties
were limited by the intrinsic dispersion in source ellipticity.
More precisely if we define $ N_{\rm eff}=\sum_i (\sigma_e/\sigma_{e,i})^2$,
the effective source number density would then be $\nbg^{\rm eff} = 25.3$
(resp. 20.7, 21.6, 21.0) arcmin$^{-2}$. This quantity represents what
could be achieved under deep space-based observing conditions.

\begin{figure}[tbh]
  \centering
  \includegraphics[width=\hsize]{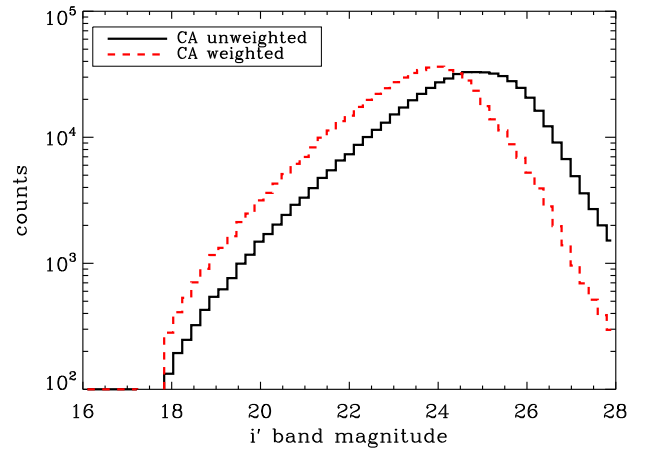}
  \caption{\small $i'$ band magnitude distribution of background sources. For clarity we represent the whole magnitude distribution although the CA catalogue only contains galaxies in the range $22<i'<26$ . The dashed red line illustrates the effective change in magnitude distribution of lensed sources when accounting for the weighting scheme of Eq. \eqref{eq:weight_scheme}. The weights are normalised so as to conserve the total number of objects.}
  \label{fig:histo-mag}
\end{figure}

\reffig{fig:histo-mag} illustrates the effect of our
weighting scheme (Eq. \ref{eq:weight_scheme}) in the magnitude
distribution of sources. The weighting scheme efficiently
down-weights objects fainter than $i'\sim 24$ and most of the signal
is carried by galaxies brighter than $i'\sim25$. Actually $i'= 24$ is
the magnitude above which shape measurement errors start to increase
above the intrinsic scatter of source ellipticities.

\subsection{Photometric redshifts}\label{ssec:data:photoz}
The number of filters available in the CFHTLS Deep images allows
a direct estimate of the redshift of each source according to its
Spectral Energy Distribution (SED) from \ugriz~bands.
We used the publicly available catalogue of photometric redshifts recently
carried out by \citet{ilbert06} for which the accuracy has been extensively
assessed against spectroscopic data\footnote{\url{http://terapix.iap.fr/article.php?id_article=586} or \url{http://cencos.oamp.fr/cencos/CFHTLS/}}.
In some cases \citeauthor{ilbert06} used 
additional non-CFHTLS data in other photometric bands to improve the
precision of their photometric redshifts. We refer the reader to this
work for further accounts. Here we briefly mention that photometric
redshifts have been checked to be reliable enough for our purpose down
to $i'\sim 25$. The authors mention that the accuracy is reduced at
redshift $\zphot>1.5$.

\begin{figure}[tbh]
  \centering
  \includegraphics[width=0.5\textwidth]{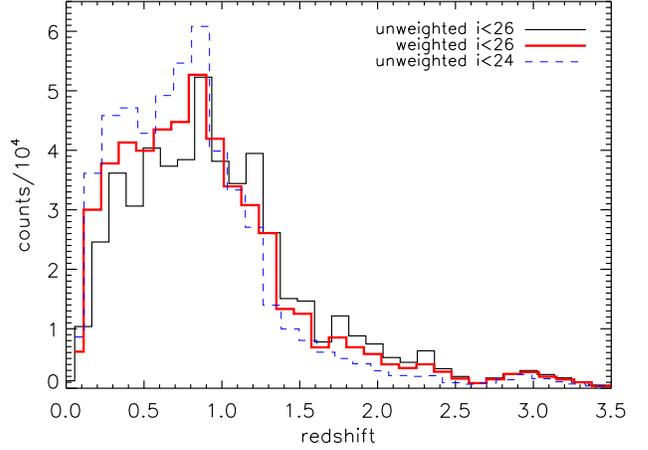}
  \caption{\small Redshift distribution of sources brighter than $i'=26$. {\it Thin Solid black:} (resp. {\it thick solid red}) the weighting scheme in Eq. \eqref{eq:weight_scheme} which reduces the contribution of  faint-distant sources is (resp. not) taken into account. For comparison the unweighted redshift distribution of $i'<24$ sources is shown {\it (dashed blue)}. This latter distribution is renormalised to the same total number of objects as the $i'\le26$ sample.}
  \label{fig:zphots}
\end{figure}

In the following we use \citet{ilbert06} photometric redshifts in three distinct ways, each of them requiring different precision on $\zphot$.
\begin{itemize}
\item To estimate the redshift distribution of the whole sample of background galaxies CA, we use photometric redshifts down to $i'\le 26$ which is a rather faint limit for accurate photometric redshifts. However the magnitude-dependent weigthing scheme of Eq. \eqref{eq:weight_scheme} means that the net contribution of lensed sources peaks at $i'\sim24$ and drops quickly for fainter objects (see \reffig{fig:histo-mag}). \reffig{fig:zphots} shows that the weighted redshift distribution of $i'\le 26$ sources is close to that of the unweighted distribution of $i'<24$ galaxies for which photometric redshifts are accurate. In addition we see on this figure that 80\% sources are at redshift below $1.5$. Therefore the relatively degraded precision of photometric redshifts beyond $z=1.5$ is not an important concern for the determination of the redshift distribution of CA. The mean weighted redshift of sources is $\overline{\zs}=0.92$ (whereas it is $\overline{\zs}=1.01$ if weights are neglected).
\item In Sect. \ref{sec:ztomo} we will need to know the redshift of background sources individually. We discuss there the effect of changing the limiting magnitude or redshift of our sample.
\item In Sect. \ref{ssec:optical-cpart}, we identify the optical counterpart of structures detected by weak lensing. We use the photometric redshift distribution of the bright galaxies present in the structure as a proxy for the structure redshift itself. In this case we will consider bright $i'<23$ lens galaxies for which photometric redshift are very accurate $\sigma_z/(1+z)\sim0.03$ down to $z\sim1.5$ \citep{ilbert06}.
\end{itemize}

\section{Mass reconstructions}\label{sec:massmap}
\subsection{Convergence maps from observed shear}\label{sec:massmap1}
From the source catalogue CA defined in Sect. \ref{sec:data}, we can
infer the shear field $\gamma(\vec{\theta})$ and deduce the associated
convergence field $\kappa(\vec{\theta})$. They are related by:
\begin{equation}\label{eq:KS93}
  \kappa(\vec{\theta}) = \int_{\mathbb{R}^2} K(\vec{\theta}-\vec{\vartheta})^* 
  \gamma(\vec{\vartheta}) \der^2 \vec{\vartheta},
\end{equation}
where $K(\vec{\theta})=\frac{1}{\pi}\frac{-1}{(\theta_1-i \theta_2)^2}$
is a complex convolution kernel \citep{kaiser93} (hereafter KS93).
The shear field is smoothed with a Gaussian filter
$G(\theta)\propto\exp(-\frac{\theta^2}{2\theta_s^2})$ with
$\theta_s=1$ arcmin. The convergence field is consequently smoothed by
the same filter. The resulting convergence maps present correlated
noise properties \citep{waerbeke00}.
\begin{equation}\label{eq:map-noise-ppties}
  \langle \kappa_n(\vec{\vartheta}) \kappa_n(\vec{\vartheta}+\vec{\theta})
  \rangle = \frac{\sigma_e^2}{4 \pi \nbg \theta_s^2}
  \exp\left(-\frac{ \vec{\theta}^2}{4\theta_s^2}\right).
\end{equation}
$\sigma_e (4\pi \nbg \theta_s^2)^{-1/2}$ characterises the noise level.
We measured a value 0.0196, 0.0225, 0.0202 and 0.0221
for D1, D2, D3 and D4, respectively.

\begin{figure*}[tbh]
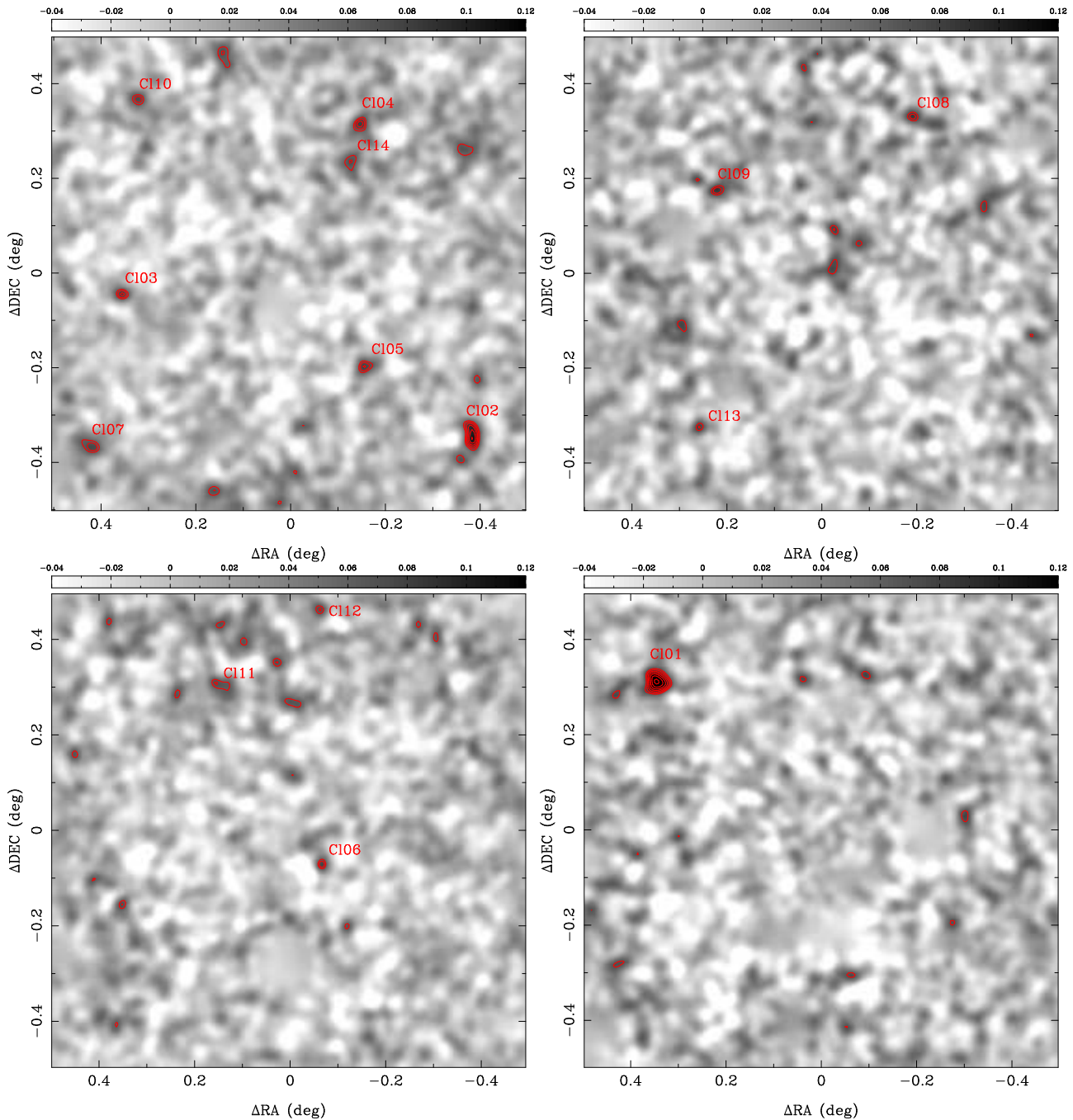

  \centering
  \includegraphics[width=8.5cm]{D1a.eps}
  \includegraphics[width=8.5cm]{D2a.eps}
  \includegraphics[width=8.5cm]{D3a.eps}
  \includegraphics[width=8.5cm]{D4a.eps}
  \caption{\small Convergence maps inferred from the ellipticity field of
    background sources for D1 (top left), D2 (top right),
    D3 (bottom left) and D4 (bottom right).
    The shear field is estimated by smoothing the
ellipticity field of sources galaxies selected in catalogue CA. Then, Eq.\eqref{eq:KS93} is used the convert $\gamma$ into the $\kappa$ field. The largest masked regions are visible as fuzzy $\kappa\sim0$ regions.
    Contours levels start at
    3$\sigma$ with a $0.5\sigma$ arithmetic increase.
    The Gaussian filtering scale is 1 arcmin. The 14 peaks with
    $\nu>3.5$ are labeled.}
  \label{fig:kappamap}
\end{figure*}

In principle, the convergence computed from Eq. \eqref{eq:KS93} must
be real and its imaginary component should only be due to noise and
possible residual systematics. We checked this assumption by rotating
the shear field by 45$^\circ$ and found the reconstructed
maps to be consistent with noise as described by
Eq. \eqref{eq:map-noise-ppties}.

The KS93 inversion in Eq. \eqref{eq:KS93} is done by a direct
summation over all sources without pixelling, smoothing and
Fourier transforming the data. This reduces boundary and
mask effects on mass reconstructions. Several techniques have been
proposed so far since the original KS93 method. Most of them are
useful in high shear regions (where $g\lesssim1$) and for small fields
of view. However the wide MegaCam images and the complex field
geometry imposed by the masks make difficult, time consuming and
unnecessary the implementation of more complex techniques. In addition,
\citet{waerbeke00} has shown that noise properties of KS93 method are
well controlled and consistent with Eq. \eqref{eq:map-noise-ppties}.

\reffig{fig:kappamap} shows the convergence maps for D1, D2, D3 and D4
deduced from the catalogue CA. Contours in units of the signal-to-noise ratio
(SNR or $\nu$) are overlaid, with $\nu$ defined as
\begin{equation}\label{eq:defSNR}
\nu = {\rm SNR} = \frac{\kappa}{\sigma_e} \sqrt{4\, \pi\, \nbg\,\theta_s^2} 
\end{equation}
In the present data we detect $\sim 46$ positive peaks with
$\nu>3$ and 5 peaks with $\nu>4$. In order to avoid too much
contamination by noise peaks but to detect as much true peaks as
possible, we therefore fix the threshold at $\nu =3.5$. The 14 peaks
detected within this limit will constitute our working sample in the
rest of the paper. We discuss in more detail the statistics
of these peaks in Sect. \ref{sec:peak-stat} and their possible association
to galaxy clusters in Sect. \ref{sec:peaks}.

\subsection{Statistics of peaks}\label{sec:peak-stat}
Several authors investigated the possibility to use convergence peaks
as clusters of galaxies candidates. Simplified analytical calculations
based on the Halo Mass Function (as inferred from the Press-Schechter
formalism for instance) provided the first predictions for wide field
imaging surveys \citep{schneider96,kruse99}. Then, thanks to the
development of numerical simulations, quantitative estimates of 
projection effects and cluster selection functions (in terms of mass
and redshift) became available
\citep{reblinsky99,jain00let,white02,padmanabhan03,hamana04,hennawi05,tang05}.

The practical implementation of a Weak Lensing Cluster Survey (WLCS) 
requires the control of noise present in observations, either due to
the intrinsic ellipticity of sources or to the intervening large scale
structures (LSS) along the line of sight. Although the
``compensated aperture mass filter'' has early
been proposed as an efficient filter for peak statistics
\citep{schneider96,kruse99,schirmer03,schirmer04,hetterscheidt05,schirmer06},
it has been shown that such a filter may not be as efficient as an
optimised filter which would account for the contribution of LSS to the noise
budget and the shape of the dark matter halos we aim at detecting
\citep{hennawi05,maturi05}. It turns out that a simple Gaussian filter of
width $\theta_s\sim1-2$ arcmin is close to the optimal linear filter and
has been extensively studied in simulations
\citep{white02,hamana04,tang05}. In addition a promising multiscale wavelets
technique has also been proposed recently \citep{starck06} but has not been
applied to real data yet. Therefore we shall use a simple Gaussian filter
with scale $\theta_s=1$ arcmin as already applied onto the mass
reconstructions of Sect. \ref{sec:massmap1}.

\begin{figure}[tbh]
  \centering
  \includegraphics[width=0.5\textwidth]{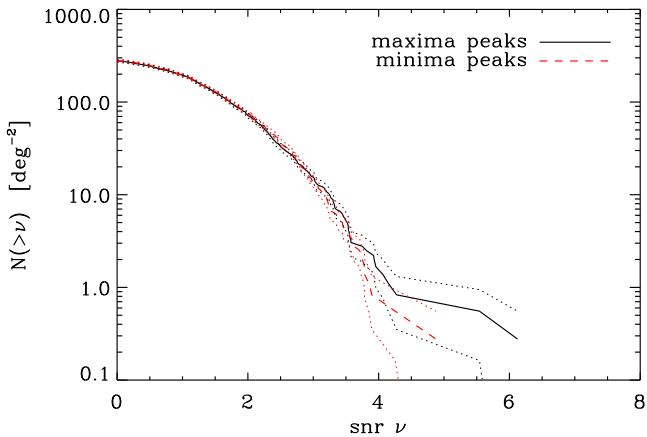}
  \caption{\small Cumulative counts of $N(>\nu)$ maxima peaks (solid black)
    and corresponding counts of $N(<-\nu)$ minima peaks (dashed blue) curve
    per square degree in the Deep survey. The surrounding dotted lines
    represent $1\sigma$ Poisson errors. The excess of positive maxima is
    marginally seen as compared to negative minima, thus showing
    the non-Gaussianity of the convergence field.
  }
  \label{fig:peaks-stat}
\end{figure}

\reffig{fig:peaks-stat} shows the cumulative number of maxima peaks
$N(>\nu)$ as well as the symmetric number of minima peaks for the four Deep
fields. The latter curve is flipped ($\nu\rightarrow-\nu$) for an easier
comparison. The net excess of maxima with $\nu \gtrsim 4$ as compared
to minima at the corresponding negative threshold is visible,
thus showing the non-Gaussian nature of the convergence field
\citep[see also][]{miyazaki02}. The statistical significance of this
excess is still low due to the large cosmic variance in such a
small sky coverage. In addition it should be kept in mind that CFHTLS
Deep fields of view were chosen to be free of any known massive nearby
cluster.

The theoretical analysis of \citet{hamana04} is well suited for a direct comparison with our results since the survey area, the smoothing scale, and the noise properties are the same. We found a satisfying agreement when considering their Fig. 7 although the sample variance is large.

An extensive study of $\kappa$-peaks statistics in the Wide CFHTLS survey
would provide valuable cosmological information for cosmic shear studies
and would help breaking some degeneracies (mainly between $\Omega_m$
and $\sigma_8$) present in the shear 2-point correlation function.
Such an analysis is beyond the scope of the present work but the full
implementation of the $\kappa$-peaks statistics in the presently released
CFHTLS-Wide data is in progress. It is noteworthy that in order to
extract the associated cosmological signal, there is no need to measure
the mass or the redshift of each peak, neither to identify them with
clusters (or projected groups, etc...).
However, instead of a blind counting exercise of convergence peaks,
we propose in the following to characterise the structures responsible
for the highest convergence maxima peaks.


\begin{figure*}[htbp]
  \centering
  \includegraphics[width=8.9cm,height=6.5cm]{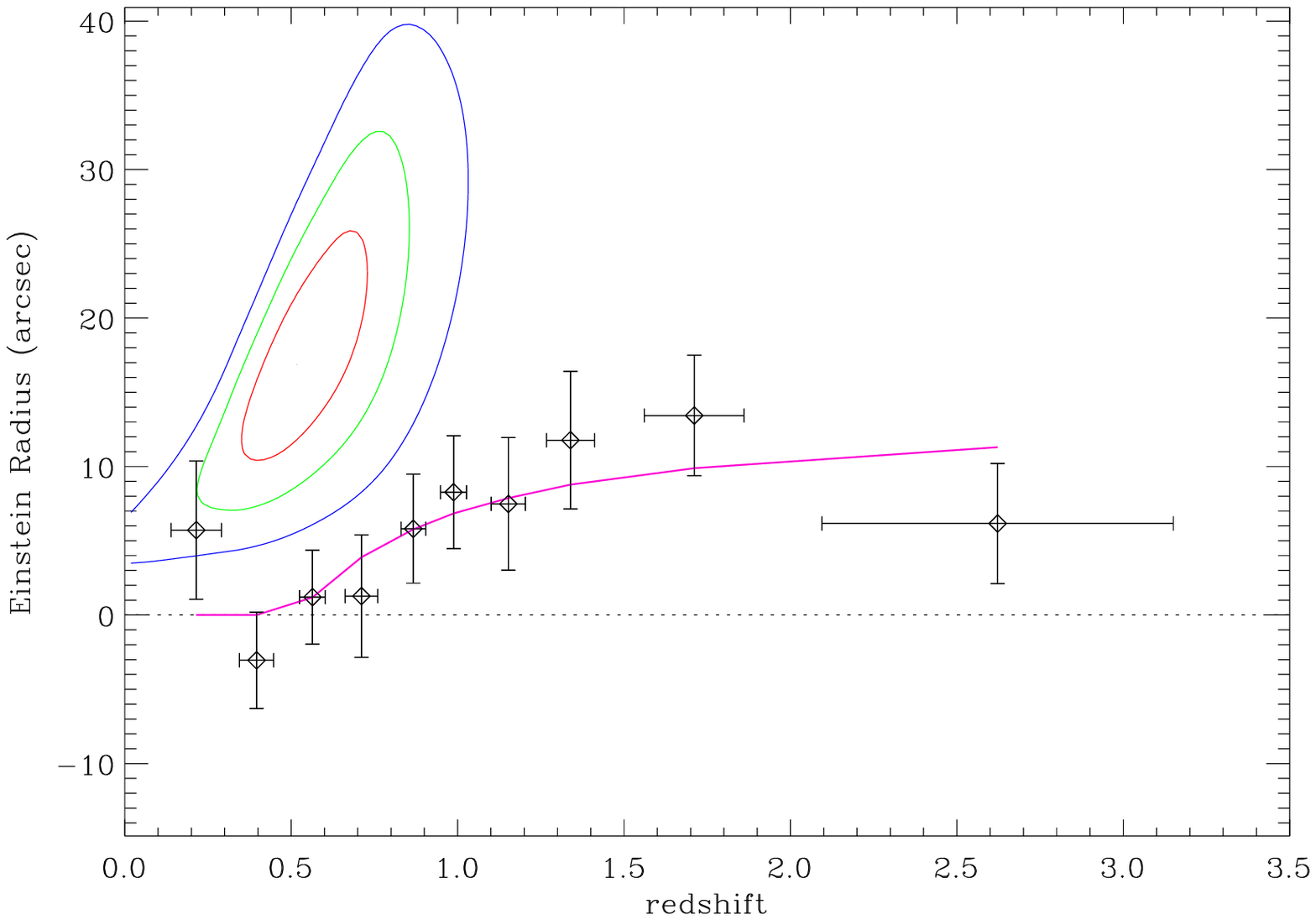}
  \includegraphics[width=8.9cm,height=6.5cm]{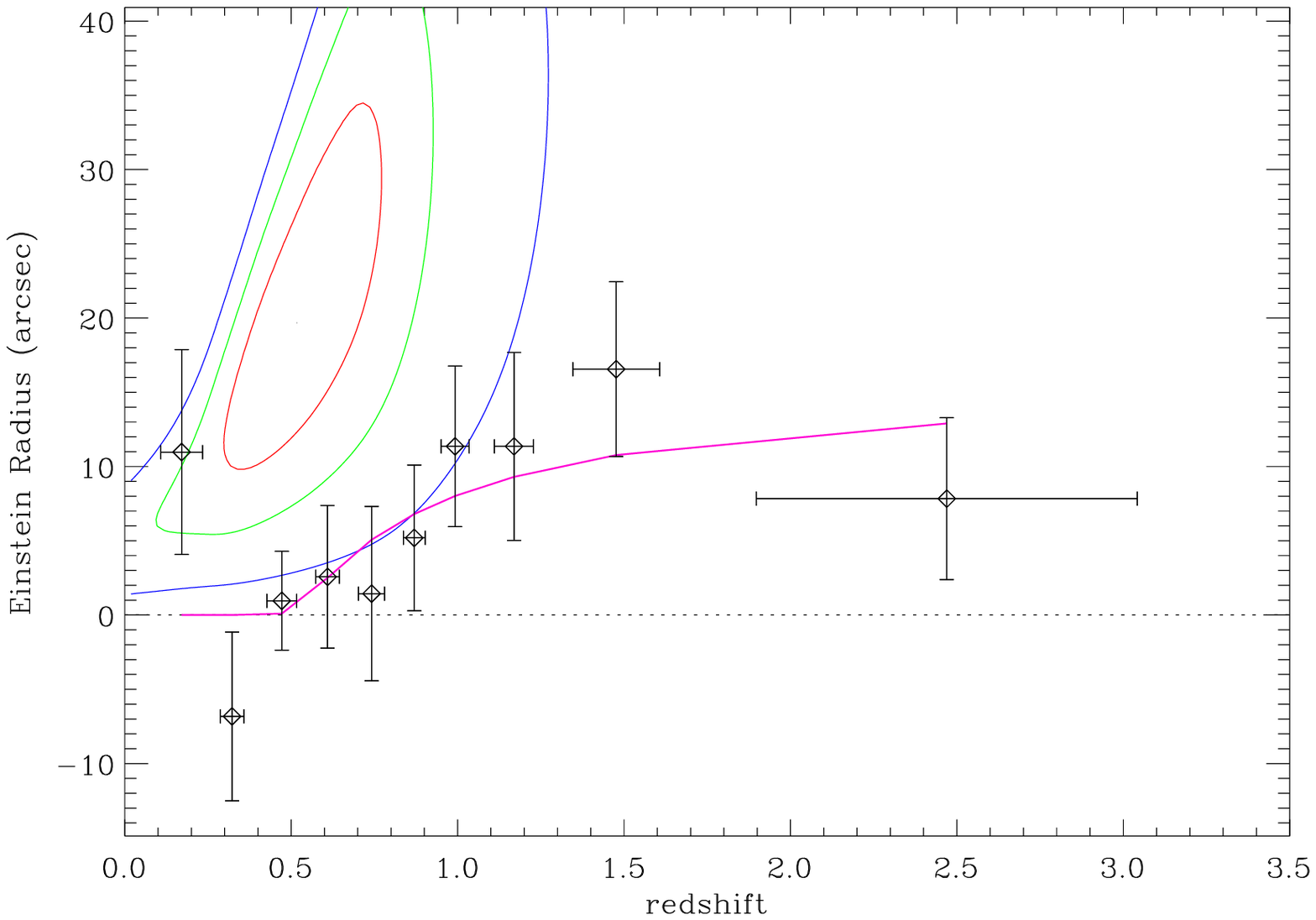}
  \caption{\small Increase of shear signal strength
      (\ie equivalent Einstein radius) as a function of source redshift
      around peak Cl-02. The source catalogue is split into 10 quantiles.
      Contours show 1,  2 and 3 $\sigma$ CL regions around the best fit
      $\rein$ and $\zl$. Peak Cl-02 exhibits the expected profile for a
      cluster with velocity dispersion $\sigma_v=760\pm110\kms$ and redshift
      $\zl=0.52\mypm{0.14}{0.11}$ with $\chi^2/{\rm dof}=0.6$ (thick line).
      The {\it left panel} uses photometric redshifts of sources down to
      $i'=26$ whereas the {\it right panel} is limited by $i'\le24.5$,
      showing that including faint sources does not affect our analysis
      in a statistically significant way.
  }
  \label{fig:tomo}
\end{figure*}

\section{Properties of $\nu > 3.5$ peaks}\label{sec:peaks}
In this section we attempt to estimate the redshift, mass and luminosity
of structure(s) responsible of the 14 $\kappa$-peaks with $\nu\ge3.5$.
This significance threshold is rather low so we expect a substantial
amount of contamination by noise fluctuations. We focus on the physical
properties of the detected peaks and test whether they can be associated
to galaxies over-densities. The first step in this peak identification
is to estimate a redshift, taking advantage of photometric redshifts.
Two methods are explored, one which is directly related to the shear
signal dependency on lens redshift and which does not require an explicit
identification of the peak with galaxies, and the other which is related
to the identification of a localised over-density in the photometric
redshift distribution of galaxies. Results from the two approaches
are summarised in Table \ref{tab:catalogue}.

\subsection{Lens tomography}\label{sec:ztomo}
The basics of lens tomography  are the following: 
in the case of a real deflector at redshift $\zl$, the shear signal must 
increase in a characteristic way as a function of the source redshift
according to the $w(\zl,\zs)=\dls/\dos$ term. Early applications can be found
in \citet{wittman01,wittman03} and \citet{hennawi05}.
The shape of the shear increase versus the source redshift allows
to derive the lens redshift. In the following, we apply this technique
to estimate the lens redshift of each peak and check whether the shear
behaviour around peaks is consistent with the lensing hypothesis or is
rather a noise fluctuation.

We measure the shear profile around each peak between 1 and 5 arc minutes from the centre, using sources having $i'\le 26$ and their individual photometric redshifts. Assuming that the lens mass distribution at redshift $\zl$ follows a Singular Isothermal Sphere (SIS) profile, the shear is simply
\begin{equation}
  \gamma(\theta,\zs) = w(\zl,\zs) \frac{\rein}{2 \theta}.
\end{equation}
The Einstein radius $\rein$ is related to the characteristic cluster
velocity dispersion by $\sigma_v= 186.2\kms \, (\rein/1 \arcsec)^{1/2}$.
We fit the shear function for the unknown lens redshift $\zl$ and 
Einstein radius $\rein$ by minimising a $\chi^2(\zl,\rein)$ of the form
\begin{equation}\label{eq:chi2tomo}
    \chi^2(\zl,\rein) = \sum_i \frac{\left( e_{t,i}-w_i
        \frac{\rein}{2\theta_i}\right)^2}{\sigma_{e,i}^2} 
 \end{equation}
where $w_i=w(\zl,z_{{\rm s},i})$, $\sigma_{e,i}$ is given by Eq.
\eqref{eq:weight_scheme} and $e_{t,i}$ is the tangential component of
ellipticity. The dependency of $\chi^2$ with $\rein$ can be easily
removed by considering that  $\frac{\partial \chi^2}{\partial \rein}=0$ 
for any $\zl$ if $\rein$ satisfies 
\begin{equation}\label{eq:thetae-estim}
   \widehat{\rein} = \frac{ 2\sum_i \frac{e_{t,i} w_i}{\theta_i\sigma_{e,i}^2} }{ \sum_i \frac{ w_i^2}{\theta_i^2\sigma_{e,i}^2}}
\end{equation}
It can be inserted into Eq. \eqref{eq:chi2tomo} to give
\begin{equation}\label{eq:chi2tomo2}
 \chi^2(\zl) = \sum_i \frac{e_{t,i}^2}{\sigma_{e,i}^2} - \frac{\left(\sum_i \frac{e_{t,i} w_i}{\theta_i\sigma_{e,i}^2}\right)^2}{\sum_i \frac{w_i^2}{\theta_i^2\sigma_{e,i}^2}}\,.
\end{equation}

In order to illustrate the method we plot on the left panel of
\reffig{fig:tomo} the value of the Einstein radius measured in the
10 $\zs$ quantiles of sources between 1 and 5 arcmin from the centre
of the peak Cl-02,
detected with $\nu=5.5$. The increase of $\rein$ with redshift $\zs$
is clear and allows an unambiguous identification of the lens redshift.
Error bars come from the scatter in observed (measurement+intrinsic)
  ellipticities as determined by Eq. \eqref{eq:weight_scheme}.
Contours show the 68.3\%, 95.4\% and 99.3\% CL regions for $\zl$ and
$\rein$. The thick curve represents the function $\rein \times w(\zl,\zs)$
for the best fit values $\zl=0.52\mypm{0.14}{0.11}$ and
$\rein=17\mypm{5.5}{4.5}\arcsec$ leading to a velocity dispersion
$\sigma_v=760\pm110\kms$. Because of the rather large statistical
uncertainties, the detailed radial shape of the shear profile
introduced in the $\chi^2$ fit is not much important. For significant
peaks in the reconstructed mass maps a redshift estimate directly
evaluated from the lensing properties of the peaks is a viable method.
It could in particular be applied to any  putative ``dark clump''
where no bright galaxies can be associated with the mass peaks.

The right panel of \reffig{fig:tomo} shows that limiting the
analysis to brighter sources (including $i'<24.5$ photometric redshifts)
does not make significant changes. We checked that this is also true for
other peaks. This suggests that, even if photometric redshift of $i'>24.5$
galaxies may suffer larger uncertainties, the effect on lens tomography, and
lensing in general, is negligible. This is due to the saturation of the curve
$\dlsdos$ at high $\zs$ and to the weighting scheme
(Eq. \ref{eq:weight_scheme}), as already mentioned in
Sect. \ref{ssec:data:photoz}. Table \ref{tab:catalogue} presents the
constraints on $\sigma_v$ and $\zl$ given by tomography for the 14 peaks.
Most of them have fitted velocity dispersion values limited
to $\sigma_v\lesssim 700\kms$. There is no massive cluster with
$\sigma_v>800 \kms$ below redshift $\sim 0.7$ in the Deep survey.
This is not a surprise as the Deep fields were initially selected
for their lack of well identified Abell clusters for example.  
In addition we observed that if the signal-to-noise ratio is too low
or tomography provides a bad $\chi^2/{\rm dof}>2$, then
the inferred lens redshift is systematically $\zl=0$. Two peaks
(Cl-04 and Cl-12) are such that tomography fails in
constraining lens redshift and velocity dispersion. However this
is may also be attributed to the fact that these two peaks are noise
fluctuations rather than being produced by real galaxy clusters.

\begin{figure*}[tbh]
  \centering
  \includegraphics[width=16cm,height=13cm]{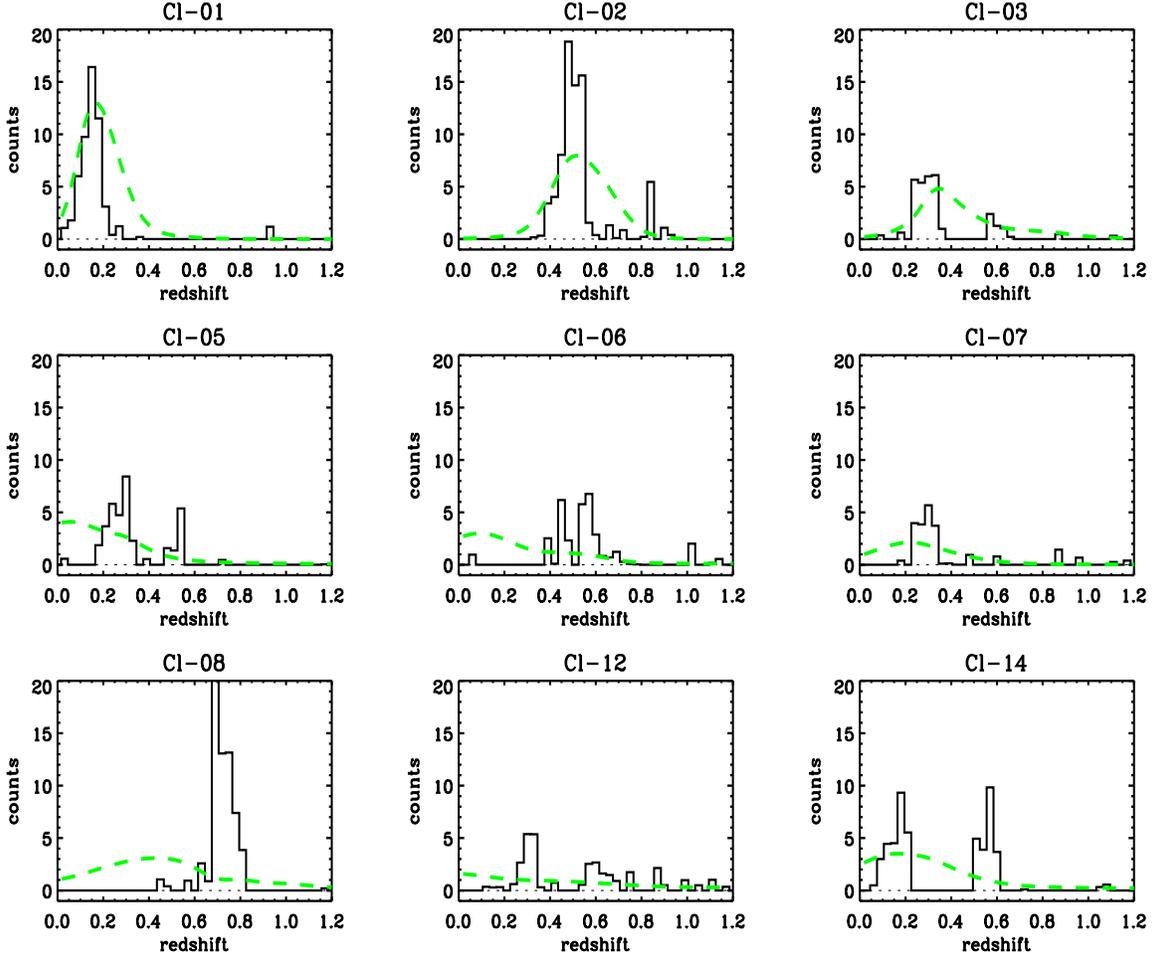}
  \caption{\small Photometric redshift distribution of galaxies around convergence peaks with a candidate optical counterpart. The green dashed curve is the lens redshift as constrained by tomography (Sect. \ref{sec:ztomo}), showing a remarquable agreement with $\zphot$ excesses for the most significant peaks. The total number of galaxies in the peaks should not be seen as richness since we did not correct for the presence of masks and the increasing incompleteness as $z$ increases.}
  \label{fig:zpdf}
\end{figure*}

\begin{table*}[htbp]
    \caption{\small Catalogue of convergence peaks / galaxy clusters in CFHTLS Deep fields.\label{tab:catalogue}}
\begin{center}
    \begin{tabular}{l | cc c cc| c ccc| cc}
      \hline\hline
      ID    & \multicolumn{5}{c|}{Convergence peak} & \multicolumn{4}{c|}{Optical counterpart(s)} & \multicolumn{2}{c}{X-ray counterpart}\\

           & $\alpha$             & $\delta$             &$\nu$&$\sigma_{v,{\rm lens}}$&         $z_{\rm tomo}$& $\Delta(\alpha,\delta)$ & $\zphot$ &$\sigma_{v,{\rm lens}}$&$\Lambda$ &$L_{X,[0.5-2.0]}$ &$T_X$\\
           & J2000                & J2000                  &   & $\kms$             &                       &  arcsec &                          &             $\kms$ &         & $10^{43}$ erg/s & keV\\ \hline
      Cl-01&$\coordra{22}{16}{58}$&$\coorddec{-17}{25}{10}$&6.2&$680\mypm{ 80}{ 70}$&$0.16\mypm{0.10}{0.08}$&( -8,  1)&$0.139\mypm{0.007}{0.007}$&$600\mypm{ 85}{105}$&$ 56\mypm{ 8}{ 8}$& $14.5\mypm{0.2}{0.2}\;^{\rm a}$ &\\
      Cl-02&$\coordra{02}{24}{27}$&$\coorddec{-04}{50}{34}$&5.5&$760\mypm{125}{105}$&$0.52\mypm{0.14}{0.11}$&( 34, 13)&$0.497\mypm{0.011}{0.011}$&$683\mypm{113}{142}$&$ 69\mypm{ 9}{ 9}$&  &  \\
      Cl-03&$\coordra{02}{27}{24}$&$\coorddec{-04}{32}{19}$&4.1&$680\mypm{110}{100}$&$0.35\mypm{0.11}{0.10}$&(  2,-11)&$0.286\mypm{0.012}{0.012}$&$611\mypm{ 88}{108}$&$27\mypm{ 6}{ 6}$& $\sim1.5\;^{\rm b}$&$1.02\mypm{0.19}{0.15}$ \\
      Cl-04&$\coordra{02}{25}{24}$&$\coorddec{-04}{10}{48}$&4.1&$460\mypm{ 90}{ 90}$&$0.00\mypm{0.20}{0.50}$&--&--&--&--& X & X \\
      Cl-05&$\coordra{02}{25}{21}$&$\coorddec{-04}{41}{33}$&4.0&$460\mypm{155}{100}$&$0.06\mypm{0.25}{0.20}$&( 27, 38)&$0.269\mypm{0.014}{0.014}$&$457\mypm{116}{159}$&$21\mypm{ 5}{ 5}$&$\sim5.2\;^{\rm c}$&$2.02\mypm{0.49}{0.28}$\\
      Cl-06&$\coordra{14}{19}{01}$&$\coorddec{+52}{36}{43}$&3.8&$490\mypm{120}{120}$&$0.09\mypm{0.18}{0.18}$&( -1,  2)&$0.533\mypm{0.025}{0.025}$&$654\mypm{132}{175}$&$19\mypm{ 5}{ 5}$&&\\
      Cl-07&$\coordra{02}{27}{40}$&$\coorddec{-04}{51}{38}$&3.8&$570\mypm{130}{120}$&$0.22\mypm{0.17}{0.16}$&(-25, 11)&$0.292\mypm{0.019}{0.019}$&$521\mypm{114}{154}$&$33\mypm{ 7}{ 7}$&$\sim6.5\;^{\rm d}$&$1.71\mypm{0.15}{0.11}$\\
      Cl-08&$\coordra{10}{01}{21}$&$\coorddec{+02}{22}{58}$&3.7&$635\mypm{215}{215}$&$0.44\mypm{0.18}{0.29}$&( -3,-27)&$0.735\mypm{0.012}{0.012}$&$416\mypm{331}{416}$&$156\mypm{13}{13}$&&\\
      Cl-09&$\coordra{09}{59}{42}$&$\coorddec{+02}{32}{20}$&3.7&$680\mypm{270}{220}$&$0.47\mypm{0.17}{0.20}$&--&--&--&--&&\\
      Cl-10&$\coordra{02}{27}{16}$&$\coorddec{-04}{07}{37}$&3.7&$480\mypm{ 90}{ 90}$&$0.16\mypm{0.14}{0.18}$&--&--&--&--&X&X\\
      Cl-11&$\coordra{14}{20}{28}$&$\coorddec{+52}{59}{22}$&3.6&$400\mypm{170}{150}$&$0.35\mypm{0.15}{0.50}$&--&--&--&--&&\\
      Cl-12&$\coordra{14}{19}{02}$&$\coorddec{+53}{08}{44}$&3.6&$330\mypm{110}{110}$&$0.00\mypm{0.36}{0.50}$&( 42,-14)&$0.296\mypm{0.023}{0.023}$&$401\mypm{149}{272}$&$35\mypm{ 7}{ 7}$&&\\
      Cl-13&$\coordra{10}{01}{30}$&$\coorddec{+01}{53}{04}$&3.6&$530\mypm{ 80}{110}$&$0.18\mypm{0.11}{0.15}$&--&--&--&--&&\\
      Cl-14&$\coordra{02}{25}{29}$&$\coorddec{-04}{15}{34}$&3.6&$460\mypm{150}{140}$&$0.16\mypm{0.50}{0.20}$&( 23, 72)&$0.153\mypm{0.011}{0.011}$&$289\mypm{144}{289}$&$ 2.0\mypm{1.4}{1.4}$&$\sim2.4\;^{\rm e}$&$1.34\mypm{0.21}{0.10}$\\
           &                      &                        &   &                    &                       &(  3,-10)&$0.569\mypm{0.024}{0.024}$&$479\mypm{184}{351}$&$25\mypm{5}{5}$&X&X\\
      \hline
    \end{tabular}\\
\end{center}
{\it Notes:} (a) ROSAT cluster
$(\coordra{22}{16}{56.2},\coorddec{-17}{25}{25.5})$ at $z=0.13$ \citep{degrandi99}, (b) XMM-LSS cluster at $z=0.31$ (XLSSC13),
(c) XMM-LSS cluster at $z=0.26$ (XLSSC25),
(d) XMM-LSS cluster at $z=0.29$ (XLSSC22),
(e) XMM-LSS cluster at $z=0.14$ (XLSSC41).
All XMM-LSS data (b,c,d,e) are from \citep{willis05,pierre06}.
Rows filled with ``--'' are likely false detections without a reliable optical counterpart. Rows filled with ``X'' are in the D1 field part of the XMM-LSS
survey but not detected in X-rays. Cl-02 is part of D1 but lies in a region
lost by the XMM-LSS survey (due to high flare rates). The second component
of Cl-14 is not detected by the XMM-LSS as it may be hidden by the foreground
$z=0.14$ component. 
\end{table*}

\subsection{Optical counterparts}\label{ssec:optical-cpart}
Here again we use photometric redshifts to check whether an optical
counterpart can be assigned to our 14 high convergence peaks with $\nu>3.5$.

We first examine galaxies in a circular aperture of 2 arcmin around
each peak. This radius corresponds to a linear physical scale of 400
to 860 kpc for a lens redshift ranging from 0.2 to 0.7 respectively.
Therefore this is the radius within which the highest
density of bright galaxies is expected, most of them being early-type
galaxies. In addition, most of these galaxies are localised in a narrow
range of photometric redshifts once the background $\zphot$-distribution
is subtracted. The background is defined in the region beyond 6 arcmin of
{\it all peaks}. In this preliminary step we only consider galaxies brighter
than $i'=23$ and with a best fit SED template type T of type earlier than
the spectral type $T=44$\footnote{The best-fit galaxy model $T$ given by
  \citet{ilbert06} therefore excludes starbusts and Irregular galaxies for
  which photometric redshifts are less reliable.}.

9 peaks out of 14 meet this criterion whereas there is no clear optical
associable counterpart for the remaining 5 peaks (rows filled with dashes
in Table \ref{tab:catalogue}). For each of the former 9 $\kappa$-peaks,
we define the cluster redshift as the location of the most prominent
$\zphot$ excess peak. We iteratively apply a 4$\sigma$ clipping
to remove outliers and get a reliable redshift and its corresponding
statistical uncertainty. Distributions can be seen in \reffig{fig:zpdf}.
The tomographic $\zl$ probability distribution
function of Sect. \ref{sec:ztomo} (marginalised over velocity dispersion)
is overlayed for comparison. We see a remarquable agreement for the most
significant peaks.

The peculiar case of Cl-14 deserves a special attention. Cl-14 clearly
exhibits a bimodal distribution that lens tomography is unable to reveal,
probably because of the low detection level. The first excess is
at $z\simeq0.15$, and is $d=75\arcsec$ away from the convergence peak location.
It is consistent with an XMM-LSS detection at a similar redshift
(see Sect. \ref{sec:xmm-lss}). However the second peak, at redshift
$z\simeq0.57$, is much closer to the convergence peaks ($d=14\arcsec$)
although it does not match an XMM-LSS peak. It is likely that both components
may contribute to the overall convergence. This is a clear example of the
projection effects already mentioned at the end of Sect. \ref{sec:lensdef}.
In the following we will consider separately these two components, labeled
Cl-14a for the $z\simeq0.15$ peak and Cl-14b for the furthest one.

We also define the luminosity-weighted optical centre using the bright
galaxies ($i'<23$) in the $\zphot$ excess range. Optical centres are
less than one arcmin away from the convergence peak location,
as expected from the spatial resolution of the mass maps. 
The observed offsets are reported in column $\Delta(\alpha,\delta)$ of
Table \ref{tab:catalogue}. From the new defined centre and the
present cluster photometric redshift, we re-estimate the lensing velocity
dispersion (central part of Table \ref{tab:catalogue}). The velocity
dispersions of Cl-14a and b are fitted simultaneously. Although they are slightly
correlated, we can deblend the system and put constraints
on both components. In any case, Cl-14b is the dominant contribution to the
convergence peak.

In order to relate lensing velocity dispersion and the excess of foreground
galaxies we measure the number $\Lambda$ of $M_r<M_0$ galaxies in a fiducial
physical radius $R_0=1.43 \Mpc$ and a slice $\delta \zphot=\pm0.10$,
with $M_0=-20$ in the rest-frame $r'$ band. The most distant of our peaks
being at $z\sim0.74$, the sample of photometric redshifts that will be close
to complete at $M_0$ corresponds to $i'\sim23.5$, which is a conservative
limit for accurate photometric redshifts. Their contribution to the
$L>L_*$ luminosity function is very low anyway. Assuming a luminosity
function with slope $\alpha$, the total luminosity in the aperture is 
\begin{equation}\label{eq:L-rich}
  L_{\rm aper} = L_* \Lambda \frac{\Gamma(2+\alpha,0)}{\Gamma(1+\alpha,q)}\simeq 1.24 \Lambda L_*\;,
\end{equation}
with $q=10^{0.4(M_*-M_0)}$ and assuming $\alpha=-1.09$, $M_*=-21.21$
\citep{blanton03}. This procedure limits the impact of bad photometric
redshifts as compared to a simple summation of galaxies luminosities.
Because photometric redshifts are imperfect, we need to subtract a background
contribution which is evaluated using galaxies outside $1.5R_0$ of
{\it all peaks} in the corresponding redshift slice. We also compensate for
the aperture area lost in masked regions. Cl-14a and Cl-14b are considered
independently.

\begin{figure}[tbh]
  \centering
  \includegraphics[width=0.5\textwidth]{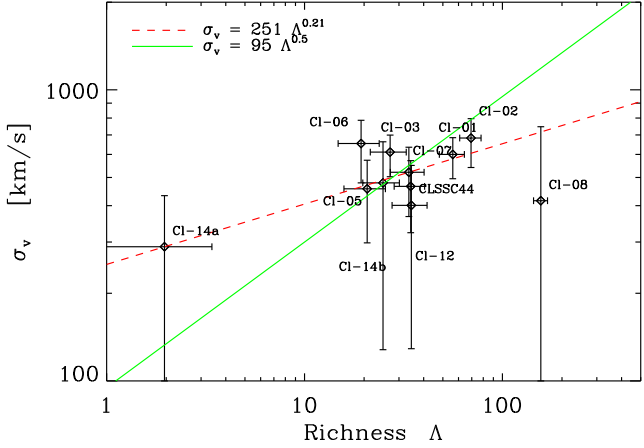}
  \caption{\small Correlation between richness and lensing velocity dispersion for Cl-01, Cl-03, Cl-05, Cl-06, Cl-07, Cl-08, Cl-12, Cl-14a-b and XLSSC44. The dotted red line is a raw linear fit leading to the scaling $\sigma_v \sim 251 \kms \Lambda^{0.21}$. If we assume a constant mass-to-light ratio, we find $\sigma_v = (95\pm 7) \Lambda^{1/2} \kms$ \ie~ $M/L_r=168\pm24 \mslsun$ ({\it solid green line}).}
  \label{fig:correl-sl}
\end{figure}

The correlation between lensing velocity dispersion and richness is shown in \reffig{fig:correl-sl} for the eight systems Cl-01, Cl-02, Cl-03, Cl-05, Cl-06, Cl-07, Cl-08, Cl-12 and the Cl-14 which is split into its two componants. We also add a peak with snr $\nu=3.4$ as it is part of the XMM-LSS cluster sample (see Sect. \ref{sec:xmm-lss}). The range of richness and velocity dispersion probed by our sample is quite narrow so it is difficult to put tight constraints on the scaling relation between these quantities. A linear regression yields : $\sigma_v \sim 251 \kms \Lambda^{0.21}$. The slope is however poorly constrained.

For our assumed isothermal mass profile, the projected mass enclosed in
radius $R_0$ is:
\begin{equation}\label{eq:M-sigR}
  M(<R_0) = \pi \sigma_v^2 R_0 / G\;.
\end{equation}
From the data we find $\sigma=96\pm7 \kms \Lambda^{1/2}$ (solid green line of \reffig{fig:correl-sl}). Combining Eq. \eqref{eq:L-rich} and Eq. \eqref{eq:M-sigR}, this translates into a $r'$ band mass-to-light ratio $M/L_r=168\pm24 \mslsun$ inside the aperture $R_0$. The $z=0.74$ peak Cl-08 slightly departs from this relation but this could partly be due evolution of galaxies luminosities. In addition, weak lensing mass estimates for such high redshift clusters are particularly sensitive to the high redshift sources. Since the accuracy of photometric redshifts is more hazardous at $\zs > 1.5$ the lensing-inferred velocity dispersion of this cluster is more uncertain. Otherwise our $M/L$ ratio is consistent with recent estimates \citep[see \eg][]{yee03,popesso06,bardeau05}. 

\subsection{Merits of lens tomography}\label{ssec:ztomo-zopt}
It is instructive to compare the redshift arising from lens tomography
($z_{\rm tomo}$) and the redshift $\zphot$ given by an excess of foreground
galaxies (\ie cluster members identified by their photometric redshift).
It is not a surprise that most WLCSs detections peak at
$\langle z_{\rm tomo}\rangle \simeq \langle z_{\rm phot} \rangle\simeq 0.25$
because is corresponds to the redshift range in which gravitational
lensing is most efficient for the source population we are
considering.

\begin{figure}[tbh]
  \centering
  \includegraphics[width=0.5\textwidth]{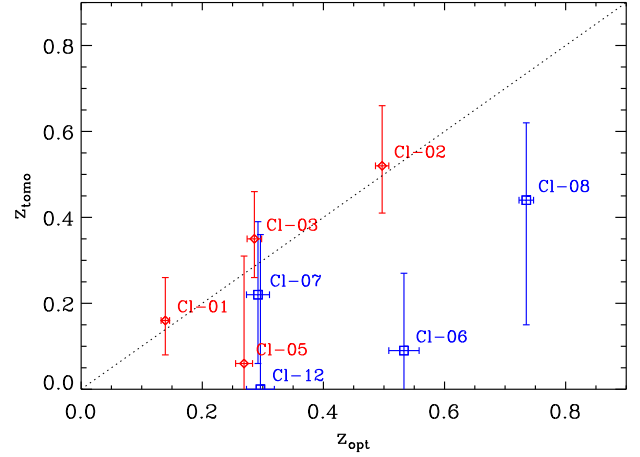}
  \caption{\small Comparison between lens tomography and direct
    photometric redshift methods to estimate the redshift of the
    $\nu>3.5$ $\kappa$-peaks. Only those peaks with a clear optical
    counterpart are included in the figure. Red diamonds represent the
    4 most significant $\nu\ge4$ peaks whereas Blue squares code for the 
    remaining 4 lower snr peaks. Due to its complexity Cl-14 is not included
    in this figure.
  }
  \label{fig:correl-z}
\end{figure}

In \reffig{fig:correl-z} we plot the comparison between these two
redshift estimates for the 8 clusters having a well identified
optical counterpart (namely clusters Cl-01, Cl-02, Cl-03, Cl-05, Cl-06
Cl-07, Cl-08, Cl-12). Cl-14 was discarded due to its apparent complexity.
Although the statistic is quite small and errors on $z_{\rm tomo}$ are large,
there is a remarquable agreement between both redshifts for the high
signal-to-noise systems, which is very encouraging. In addition we already
pointed out that Cl-04, for which tomography fails in giving a
lens redshift, has no optical (nor X-ray) counterpart and is likely a
false detection. Tomography helps eliminating such cases.

In order to improve this correlation, it will be important to improve
the quality of photometric redshifts, especially for the faintest background
sources. It is not clear how much the catastrophic redshifts in the source
catalogue perturb lens tomography, but certainly for accurate tomographic
redshift estimates it will be important to increase the number of filters
for photometric redshifts, especially in infrared bands \citep{bolzonella00}.
In addition, it is noteworthy that our sample is made of low mass clusters
($\sigma_v\sim400-600 \kms$). For more massive clusters, like those expected
in the CFHTLS-Wide survey, the method will greatly gain in accuracy and
reliability \citep{hennawi05}.

\subsection{Comparison to X-ray data}\label{sec:xmm-lss}
The CFHTLS-Deep D1 field is part of the XMM-LSS survey\footnote{
  \url{http://vela.astro.ulg.ac.be/themes/spatial/xmm/LSS/index_e.html}, 
  see also \url{http://l3sdb.in2p3.fr:8080/l3sdb/}}.
We therefore took advantage of the publicly available X-ray database
to cross-correlate our sample of $\kappa$-peaks/clusters with those
X-ray detections published in \citep{valtchanov04,willis05,pierre06}. 

The matching is very good. Over the seven $\nu>3.5$ peaks detected in D1,
four are XMM-LSS clusters with luminosity $1.5\
10^{43}\le L_{\rm X,bol}\le6.5\ 10^{43}$ erg/s and temperature $1\lesssim
T_X\lesssim 2$ keV. However the most pronounced $\nu=5.5$ D1 peak,
namely Cl-02, is not part of the XMM-LSS sample because it falls in
a region lost by the X-ray survey \citep[pointing G12,][]{pierre06}.
Peaks Cl-04 and Cl-10 do not exhibit any optical counterpart and are not
detected in X-ray either. This confirms that Cl-04 and Cl-10 are likely 
false detections due to noise fluctuations.

There are 9 publicly available XMM-LSS clusters in the D1 field of
view in the classes C1-C2\footnote{4 additional C3-class clusters could
  be added to this list but their detection and physical properties are
  more hazardous}.
4 of them are part of our weak lensing cluster sample although we
note that a $\nu=3.4$ peak at $\alpha=\coordra{02}{24}{31.7}$ and
$\delta=\coorddec{-04}{13}{55}$ is also part of the XMM-LSS sample
(XLSSC44 at $z=0.26$, $T_X=1.37\mypm{0.28}{0.16}$ keV). This cluster
has been missed by the weak lensing survey because it does not meet
the $\nu>3.5$ detection threshold. Its lensing velocity dispersion is
$\sigma_{v,{\rm} lens}=466\mypm{105}{143}\kms$. The remaining 4 XMM-LSS
clusters which are not part of our sample: XLSSC38 ($z=0.58$, $T_X$
unknown), XLSSC11 ($z=0.05$, $T_X=0.64\mypm{0.11}{0.07}$ keV), XLSSC29
($z=1.05$, $T_X=4.07\mypm{1.72}{0.99}$ keV) and XLSSC5 ($z=1.05$,
$T_X=3.67\mypm{3.50}{1.33}$ keV) are either very low or very high
redshift clusters for which lensing efficiency is low. Therefore
it is not surprising that they are missing in our weak lensing sample.

Note also that the peak Cl-01, located in D4 and which is the strongest
peak matches an X-ray detected ROSAT cluster at redshift $z=0.13$
and luminosity $L_{X,[0.5-2.0]}=14.5\pm2.5\times10^{43}$ erg
s$^{-1}$.
\citep{degrandi99}.

\begin{figure}[tbh]
  \centering
  \includegraphics[width=\hsize]{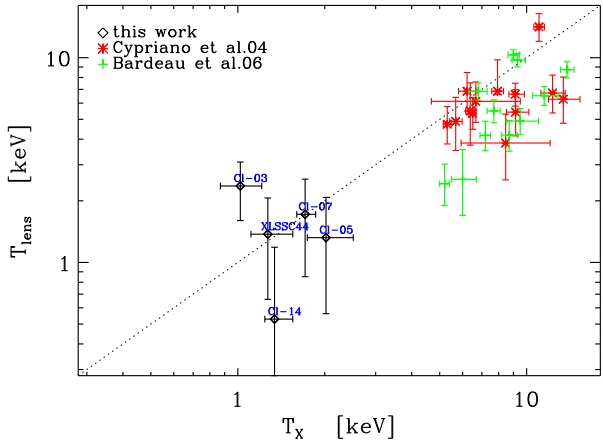}
  \caption{\small  Correlation between X-ray and lensing-inferred
    temperature $k_B T_{\rm lens}= \mu m_H \sigma_v^2$ (black diamonds)
    for Cl-03, Cl-05, Cl-07, Cl-14 and XLSSC44. The results of
    \citet{cypriano04} on massive clusters are reported for comparison
    (red stars) as well as those of \citet{bardeau06} (green crosses).
    The bisectrix line (dotted) represents energy equipartition between
    hot ICM gas and dark matter + galaxies collisionless fluid.}
  \label{fig:correl-sx}
\end{figure}

We now compare lensing velocity dispersion and X-ray temperature for
the five D1 XMM-LSS clusters Cl-03, Cl-05, Cl-07, Cl-14 and XLSSC44.
More precisely and following \citet{cypriano04}, we compare the lensing
velocity dispersion inferred temperature $k_B T_{\rm lens}= \mu m_H\sigma_v^2$
of dark matter particles to the X-ray temperature of hot ICM gas particles.
$\mu m_H$ is the mean particule weight per free electron. It is set to the
same value $\mu=0.61$ as \citet{cypriano04}. We also include in
the comparison data from the study of \citet{bardeau06} of 11 X-ray
luminous clusters at $z=0.2$. Under the assumption of energy equipartition
these temperatures would be equal. If non gravitational sources of gas
heating/cooling are at work we expect some departure from this
relation $T_X \ne T_{\rm lens}$. Conversely the mass (and thus
$\sigma_v$ and $T_{\rm lens}$) of shear-selected clusters may be
increased by projections of unrelated material along the line of sight.

Although the statistics is rather poor, \reffig{fig:correl-sx}
suggests that shear-selected clusters are well aligned onto the bisectrix
$T_X = T_{\rm lens}$. Gas and collisionless
particles share the same amount of energy. This behaviour seems to be
less true for massive clusters. For example \citet{cypriano04}
found that for $T_X \gtrsim 8$ keV, the gas is hotter than expected by
pure gravitational effects. This supports the presence of off-equilibrium
physical processes (unrelaxed clusters, merging).

The low redshift component of Cl-14 (Cl-14a) has been used for
\reffig{fig:correl-sx} although it should be considered with caution.
Both lensing and X-ray properties of this cluster may be
contaminated by projection effects: Cl-14b seems to dominate the lensing
signal whereas the nearby Cl-14a dominates the X-rays signal.

\section{Discussion and conclusion}\label{sec:conclu}

In this work we attempted to analyse the weak lensing signal in the 4
deg$^2$ images of the CFHTLS Deep fields. 
For a proper signal extraction we payed special attention
to the removal of residual systematics and got a large sample of
distant lensed sources, thanks to the exceptional depth of the images.
We then used standard KS93 inversion technique to infer the
projected surface mass density (\ie the convergence field) and
focused on maxima peaks with a signal-to-noise ratio $\nu>3.5$. We
found 14 such peaks and discussed the possibility to use the
statistics of maxima peaks as a test for the non-Gaussianity of the
convergence field. The lack of massive clusters is not surprising since
the Deep field of view is only 4 deg$^2$ and the 4 MegaCam fields were
selected {\em a priori} as free from already known rich clusters.
It turns out that half of the $\nu>3.5$ convergence peaks
(either with or without an optical counterpart) are in D1 and at redshift
$z\sim0.28$, \ie in the redshift range with best lensing efficiency.
How much of this excess is due to the 10\% lower noise level in D1 relative
to D2, D3, D4 or is pure sample variance (enhanced by the strong spatial
clustering of galaxy clusters)? The latter is our favoured explanation
since a 10\% change in SNR for D1 peaks would not significantly change
the ranking of peaks in Table \ref{tab:catalogue}. At $z\sim 0.28$,
$1\deg$ scale corresponds to $15 h_{70}^{-1}$ Mpc and the cosmic
variance of the cluster-cluster correlation function is still
high.

Looking for excesses of galaxies around convergence peaks we found that $\sim35$\% of our $\nu>3.5$ candidates turn out to be false or inconclusive detections. Our cluster candidates are not very massive systems but look more like small clusters / large groups having $400\lesssim\sigma_v \lesssim 600 \kms$. Most of them lie at redshift $\sim 0.3$ although we found a clear detection at $z=0.74$. All the D1 XMM-LSS clusters that lie in the lensing relevant redshift range $0.1\lesssim z \lesssim 0.6$ are detected with a SNR $\nu \gtrsim 3.4$.
The completeness of WLCSs is however lower than X-ray
techniques for clusters detections. If one aims at reducing the amount
of false detections (higher efficiency), the sample completeness
of WLCSs turns out to be very low. This has already been pointed out
in simulations \citep[\eg][]{white02,hamana04,hennawi05}.

Projections effects were observed in at least one of our detections (Cl-14). The other systems turn out to have lensing-inferred mass (or velocity dispersion) properties consistent with their optical ($M/L_r\sim170\mslsun$) and X-rays ($\sigma_v^2\propto T_{\rm lens}\simeq T_X$) counterparts.

We used lens tomography around $\kappa$-peaks to estimate the deflector redshift as well as its velocity dispersion. To this end we made an extensive use of photometric redshifts in the sample of background sources. With accurate photometric redshifts, tomography can improve weak lensing cluster surveys capabilities since it helps distinguishing real clusters and noise fluctuations. The agreement between tomographic redshifts and photometric redshifts of cluster members is remarquable for the most significant peaks.

Although the field area of the CFHTLS Deep survey is not wide enough
for cosmological application, we have demonstrated that CFHTLS image
quality is well suited for WLCSs as it is the case for cosmic shear
signal extraction \citep{semboloni06}. The full implementation of this
technique to the CFHTLS Wide survey is on-going and will provide us
with a few hundred shear-selected clusters. The large sky coverage
will balance the lower density of background sources as compared to
the Deep fields and will clearly favour the detection of higher mass
systems with velocity dispersion in the range $700-1200\kms$. 
Given our findings for the Deep fields, we can forecast that
a detection threshold as high as $\nu \gtrsim 4.5$ will be required
for a robust WLCS in the shallower Wide survey.
This will be done at the expense of finding
intermediate mass structures with $\sigma_v \sim 400-700 \kms$.

In addition half of this survey (the W1 region) will also be covered
by the XMM-LSS survey. This will give the necessary calibration of
scaling relations between mass and direct observables for clusters of
galaxies to be used as efficient cosmological probes. A more detailed
comparison with the performances of ongoing other clusters survey
techniques (optical, SZ with Planck) will also become possible.
The low completeness of WLCSs is balanced by their well controlled
selection function since one needs cosmological simulations
with relatively low resolution and essentially no gas physics.

The opposite approach is also possible. We mentioned
in Sect. \ref{sec:peak-stat} that it would be more interesting
to use the statistics of $\kappa$-peaks as a test of the non Gaussianity
of the convergence field. This contains complementary information
on cosmological parameters relative to the cosmic shear 2-point
correlation function. Like \eg the skewness it helps breaking the
observed degeneracy between $\Omega_m$ and $\sigma_8$ with shear
correlation functions. In this respect it is not necessary to check
whether individual peaks are real or false detections nor to assign
a redshift and mass with expensive follow-up for each convergence peak.

Both WLCSs (with cluster identifications) and raw $\kappa$-peaks
statistics are complementary applications of weak gravitational
lensing. They both will soon provide new insightful constraints
on the evolution of large-scale structure driven by Dark Matter and
perhaps giving important clues on the behaviour of Dark Energy as a
function of redshift.

\begin{acknowledgements}
  We are thankful to R. Pell\'o and F. Ienna for making photometric
  redshifts available and for useful comments on their handling.
  The comments of the anonymous referee greatly helped improving the
  quality of this work. We acknowledge fruitful discussions
  with C. Benoist, A. Blanchard, C. Marmo, Y. Mellier and L. Olsen.
  RG is supported at LATT by a postdoctoral contract \#1019
  from the CNRS. We thank the Programme National
  de Cosmologie of the CNRS for financial support.
\end{acknowledgements}

\bibliography{references}

\end{document}